\documentclass[a4paper]{article}   

\usepackage{amsmath}
\usepackage{mathptmx}						
\usepackage[english]{babel}
\usepackage[T1]{fontenc}
\usepackage{hyperref} 
\usepackage[pdftex]{graphicx}
\usepackage{cite}					
\usepackage{ifpdf}
\usepackage{color}
\usepackage{inputenc}
\usepackage{authblk}

\begin{document} 

\title{High-sensitive Optical Pulse-Shape Characterization using a Beating-Contrast-Measurement Technique}

\author[1]{V.~Roncin*,~S. Fève,~A. Millaud, ~R. Cramer,~J-C.~Simon}
\author[2]{and Y. Jaouën}

\affil[1]{CNRS UMR 6082 Foton, Enssat 6 rue de Kerampont, F-22300 Lannion Cedex, France}
\affil[2]{Institut TELECOM/Telecom Paris Tech, 46, rue Barrault, 75634 Paris, France}
\date{}

\thanks{Corresponding author information \\
E-mail: \textit{vincent.roncin@enssat.fr}, \\
Telephone: +33 (0)2 96 46 91 50}

\maketitle

\begin {abstract}
Ultrahigh-speed optical transmission technology, such as optical time domain multiplexing or optical signal processing is a key point for increasing the communication capacity. The system performances are strongly related to pulse properties. We present an original method dedicated to short pulse-shape characterization with high repetition rate using  standard optical telecommunications equipments. Its principle is based on temporal measurement of the contrast produced by the beating of two delayed optical pulses in a high-bandwidth photodetector. This technique returns firstly reliable informations on the pulse-shape, such as pulsewidth, shape and pedestal. Simulation and experimental results evaluate the high-sensitivity and the high-resolution of the technique allowing the measurement of pulse extinction ratio up to 20~dB with typical timing resolution of about 100~fs. The compatibility of the technique with high repetition rate pulse measurement offers an efficient tool for short pulse analysis.
\end {abstract}

\maketitle 

\section{Introduction}
Optical short-pulse characterization is a fundamental goal in physics for short phenomena characterization \cite{bucksbaumScience2007} as well in the area of high capacity optical communication engineering based on Optical Time Domain Multiplexing (OTDM) technology~\cite{WeberJLT2006}.
In this last application, the pulse train which has to carry out optical data is build from interleaving of several optical short pulse train delayed  in order to increase the repetition rate.
The most important points in optical pulse characterization are the pulsewidth, commonly defined as the full width at half maximum (FWHM), shape and pedestal. Particularly in the OTDM context, significant pulse pedestals can lead, to great instability caused by inter-pulse interference ~\cite{SaruwatariSTQE2000}. Specific solutions have been proposed for pulse-shape control in order to measure extinction ration improvement\cite{SmithOL1990} or pulse pedestal reduction\cite{GossetPTL2001}.
On the other hand different techniques for optical pulse characterization have been developed for the shape monitoring, the pulse energy and the effects induced by its interaction with material.
Non-linear autocorrelators~\cite{SalaJQE1980} which are commonly used for ultra-short pulse measurement, are not able neither to give real pulse-shape nor to evaluate precisely the pulse duration because of the shape hypothesis made for the deconvolution. However the temporal pulsewidth is obtained with a low precision of about 20\% if the initial hypothesis is a Gaussian or a hyperbolic secant shape. Note that the obtained trace obtained is necessary symmetrical which is leading to a lack of information on the pulse pedestal. 
Other methods for time analysis with high repetition rate has been proposed in the 90's years such as streak camera~\cite{PradeOC1994} and optical sampling~\cite{OhtaECOC1998}. Different approaches have been investigated as all optical sampling using SOA~\cite{JiangJQE2001} or fiber-based parametric amplification~\cite{LiPTL2001}. Optical sampling with repetition rate up 640~GHz~\cite{YamadaPTL2004} has been obtained. 
Combining time and frequency resolved measurement, many interesting techniques have been investigated allowing to give instantaneous phase information as well-known Frequency resolved Optical Gating (FROG)~\cite{TrebinoJOSA1993} and other derived methods~\cite{WalmsleyJOSAB1996},~\cite{DorrerOL2002},~\cite{GossetJLT2006}. Based on spectrum modulation and Fourier analysis, these different approaches are extensively studied for pulsed source shaping~\cite{ShankAP1982},~\cite{NibberingJOSAB1996} and for chirp management in optical communication~\cite{SuzukiJLT1993}.  

We present in this article an extensive analysis of our original approach for pulse characterization based on homodyne Beating Contrast Measurement (BCM) technique. This method required only low-bandwidth photodetector, compared to pulse bandwidth and a conventional electrical sampling oscilloscope~\cite{RoncinECOC2003}. Two pulse trains from a same laser source are decorrelated using optical fiber section in order to avoid optical interference, precisely delayed and afterwards recombined in the photodetector. By controlling the delay between the pulses and measuring the contrast of the beating detected we realize the pulse-shape measurement. The theoritical limitations of the BCM technique have been investigated through numerical modelling. Experimental results allow the validation of the measurement technique using pulse-shape characterization of two different pulse sources. The first one is based on electrical commutation in semiconductor laser (Q-Switched laser) producing weakly coherent laser emission with highly chirped pulses. The other source is based on fiber-doped-mode-locked laser (Mode-Lock laser) producing greater coherent pulse emission with low chirp.

\section{Principles of BCM technique}
\begin{figure}
\centering
\includegraphics[scale=0.7]{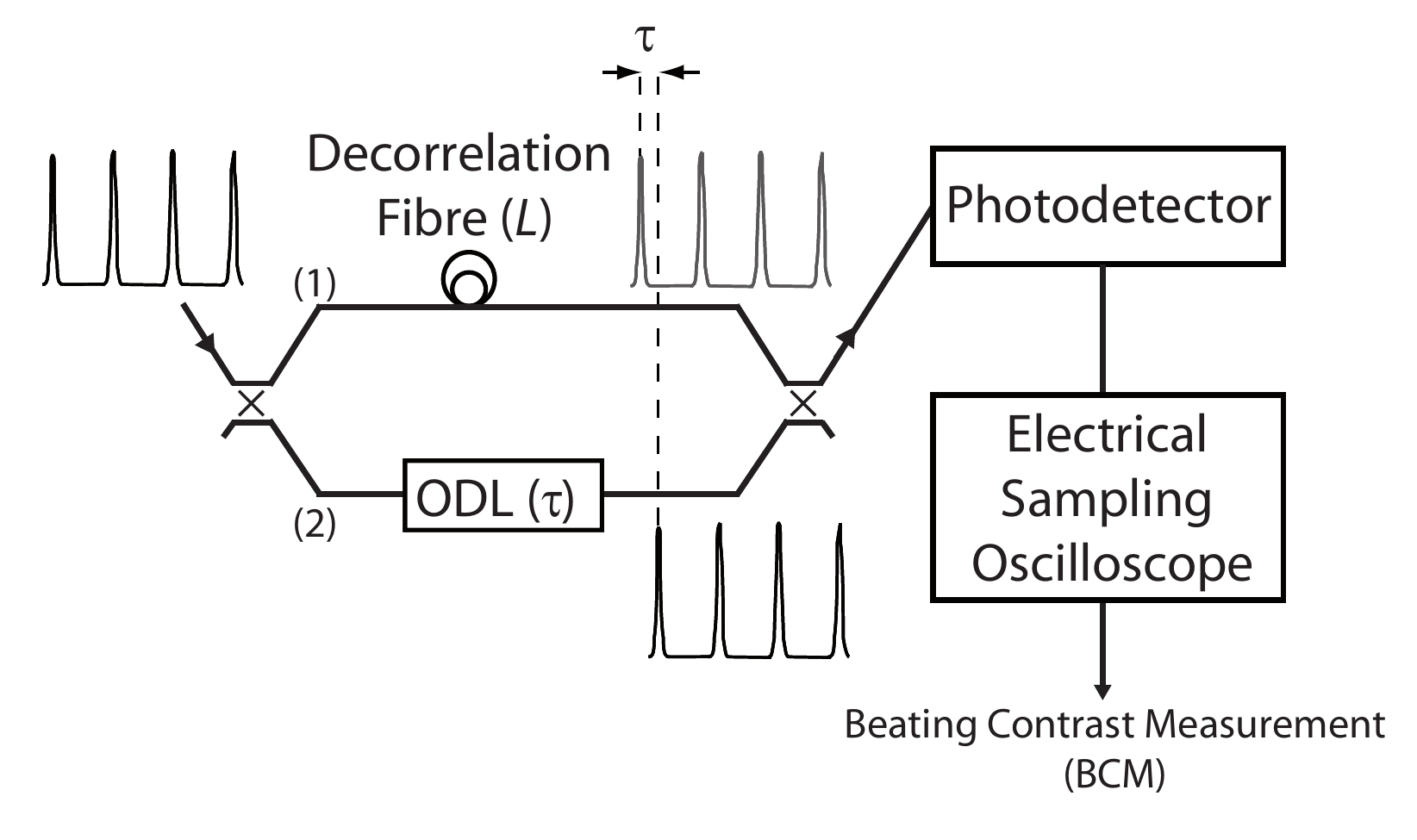}
\caption{Principle of the Beating Contrast Measurement (BCM) technique. The set-up is composed of a variable Optical Delay Line (ODL) in order to delay the pulse train in path (2) as compared to the path (1), a photodetector and an electrical sampling oscilloscope with a bandwidth greater than the pulse train repetition frequency $N$.}
\label{fig.figure1}
\end{figure}

The method is described thanks to the set-up presented in Figure~\ref{fig.figure1}. The incoming pulse train is separated onto the arms of the fiber-based Mach-Zehnder interferometer. The field in the path (1) is decorrelated as compared to the field in the path (2) assuming that the decorrelation fiber section is designed in order to avoid interference in the output 3~dB coupler. This operation is fundamental in the technique because the two optical pulses in interaction in the photodetector has to be decorrelated in order to avoid a coherent beating and to cover all the values of phase difference between the pulses and then to obtain the entire contrast of the beating.
This point is critical because the technique is based, as it will be described analytically in the following of the article, on the measurement of the beating contrast. thus, the instantaneous phase-difference between the pulses has to be a random parameter which only depends on the phase fluctuations induced by the interferometer environment. It is the reason why, the absolute state of phase of the pulses in interaction in the detector must be different. 
So, using an electrical sampling oscilloscope it is possible to observe the accumulation of all the values of the electrical phase-difference in the range [0,2$\pi$]. For a long time acquisition using histogram measurement, the whole description of beating amplitude contrast is possible. A photodetector with a cut-off frequency greater than the repetition frequency of the pulse train is used to describe the beating contrast for each value of the delay between the two pulses. 

\begin{figure}[!ht]
\centering
\includegraphics[scale=1.8]{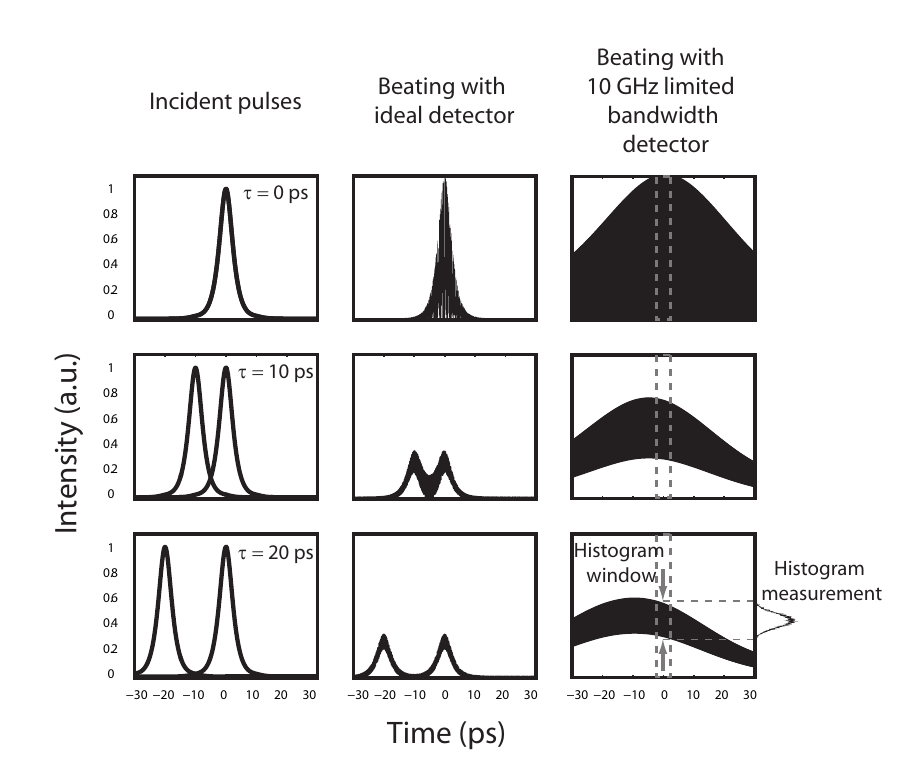}
\caption{Simulation of the BCM technique for different values of the delay $\tau$ between the two successive pulses. On the left figures, we show the incident pulses. On the middle figures, we show the beating of the pulses in an ideal photodetector with an infinite bandwidth. On the right figure, we show the beating in a 10~GHz bandwidth photodetector. We also observe the histogram window allowing the histogram measurement in order to measure the contrast of the beating.}
\label{fig.figure2}
\end{figure}

In order to make a rigorous description of the BCM technique, we present in Figure~\ref{fig.figure2} a simulation of its principle. According to the set-up previously described, we introduced a pulse-train of 10~ps pulsewidth (left column). The beating is firstly shown at the output of the interferometer using an infinite (ideal) bandwidth detector (center column). Finally, the beating is observed using a 10~GHz detector (right column) directly plugged on a high bandwidth sampling oscilloscope. This simulation is presented for three different delays: $\tau\,$=~20, 10 and 0~ps. For these different delays we show that closer are the pulses and higher is the contrast of the beating. So, it is then possible to measure the contrast of the beating for different delays between the two separated pulses~($\tau$). 
Experimentally, we measure the beating contrast in a histogram window using the histogram function of the sampling oscilloscope. The simulation which takes into account the photodetector bandwidth limitation is clearly shown on the curves on the right of Figure~\ref{fig.figure2}. 
By measuring the minimum and maximum value of the beating, it is possible to determine the contrast. Finally, we deduce thanks to the appropriate hypothesis, the shape of the pulse versus the delay between the pulses. In the following part, we propose an analytical description of the technique, in order to study the performances and the limitations of the BCM technique.

\subsection{Phase description}

In this part, we introduce the formalisme used hereafter for the optical phase description in the modelisation approach. We also suggest a discussion about the validity of the analytical description if the phase of the pulse is linear or nonlinear. We define an expression of the phase of the optical pulse by the introduction of the chirp contribution:

\begin{equation*}
\begin{split}
\phi _{pulse}(t) = \phi_{init}+\phi_{chirp}(t)
\end{split}
\end{equation*}

Where $\phi_{init}$ is the initial phase of the pulse defined arbitrarily at the pulse beginning. And $\phi_{chirp}$ is the instantaneous phase in the pulse which depends on the type of the process of generation of the optical pulse. For example, a mode-locking or a gain-switching process are not producing the same phase evolution in the pulse. Then, the chirp may be consider as linear or non-linear.

The decorrelation process is described with the delay time $T$ which is given for $T=n*L/c$, with $c$ the velocity of light in vacuum, $n$ the propagation index and $L$ the length of the decorrelation fiber. This process imposes that, for a delay higher than the optical source coherency time, $\phi_{init-1}$ for the path (1) and $\phi_{init-2}$ for the path (2), are supposed independent and correspond to a random phase. 

Then, in the BCM technique, the measurement is achieved thanks to the variation of the contrast produced by the beating of the two pulses in the photodetector. Then, the discrete changing of the delay between the pulses is defined by $\tau$. Then, one of the two pulses is taken as a reference and the other is shifted by $\tau$ as compared to the reference in order to obtain the expression of the phase for each pulses. Moreover, the chirp expression is identical for each pulses. We give the expression of the phase of the pulses (1) and (2):

\begin{equation*}
\left\{\begin{matrix}
\phi _{pulse-1}(t) = \phi_{init-1} + \phi_{Chirp}(t)
\\
\phi _{pulse-2}(t+\tau) = \phi_{init-2}+ \phi_{Chirp}(t+\tau) + \phi_{T}
\end{matrix}\right.
\end{equation*}  

Therefore, we will consider in the nest part the expression of the beating in the photodetector. So, we introduce the phase difference between the pulses during the measurement process.
In order to study the impact of the phase on the beating expression we suggest two hypothesis for the chirp expression:

\begin{itemize}
		\item{The chirp is linear and its expression is simply obtain thanks to a constant $C$ corresponding to the chirp evolution into the pulse as well:$\phi_{Chirp-L}(t) = C*t$}
		\item{The chirp is nonlinear and its analytical expression $\phi_{Chirp-NL}(t)$ is not the purpose of this article}
\end{itemize}
			
The phase difference is then suggested for these two previous hypothesis.

- For a linear chirp it is possible to express the phases as following:

\begin{equation*}
\left\{\begin{matrix}
\phi _{pulse-1}(t) = \phi_{init-1} + C*t
\\
\phi _{pulse-2}(t+\tau) = \phi_{init-2} + C*(t+\tau) + \phi_{T} = \phi_{init-2} + C*t +C*\tau + \phi_{T}
\end{matrix}\right.
\end{equation*}  	

We obtain the expression of the phase difference:

\begin{equation*}
\Delta{\phi(\tau)}= \phi _{pulse}^{(2)}(t+\tau) - \phi _{pulse}^{(1)}(t) = \Delta{\phi_{0}} + C*\tau
\end{equation*}

With $\Delta{\phi_{0}} = \phi_{init-2} - \phi_{init-1} + \phi_{T}$ that may be considered as independent to $t$ and $\tau$. For a correct decorrelation operation, its value is randomized and moving slowly as regard to eventual fluctuations introduced in the experimental set-up.

- For a nonlinear chirp, the expression of the phases is more complicated and depends on the pulse generation process:

\begin{equation*}
\Delta{\phi(\tau)} = \Delta{\phi_{0}} + \Delta{\phi_{chirp-NL}}(\tau)
\end{equation*}

With $\Delta{\phi_{chirp-NL}}$ is a nonlinear fonction which depends on the delay $\tau$.

Considering these expressions, we will base further hypothesis of the measurement analytical description, on the linearity or not of the pulse phase.

\subsection{Analytical description}
$E_1(t,0)$ is the reference pulse train which is propagated in the optical path (1) and $E_2(t,\tau)$ the delayed pulse train in the optical path 2. The beating of the two optical pulses is produced into the photodetector. The time reference of the beating is given by the pulse field $E_1$ which is not delayed ($\tau=0$). We note that the decorrelation is not introduced in the analytical model because the delay is fix for the measurement. The intensity of the beating into the photodetector is then given by the expression (\ref{Eq.Equation1}).

\begin{equation}
\begin{split}
I(t,\tau)= &\, E_1(t) * E_{1*}(t) + E_2(t+\tau)*E_{2*}(t+\tau)\\
           &\; + 2 \sqrt{ E_1(t)*E_{1*}(t)*E_2(t+\tau)*E_{2*}(t+\tau}) \, \cos\left[ \Delta{\phi(\tau)}\right]
\end{split}
\label{Eq.Equation1}
\end{equation}


Several hypothesis may be considered:\\ 

- The contrast measurement is complete as soon as $\Delta{\phi(\tau)}$ has described the range [0,$2\pi$].  \\
- The limited photodetector bandwidth modifies the electrical response of the beating. In order to get the more sensitive measurement, the intensity of the contrast has to correspond to the beating between only two pulses. Consequently, for an optical pulse train of~$N$~GHz, the bandwidth of the detector must be close to the repetition rate of the pulse train. For a bandwidth lower than $N$, the analytical expression of the beating intensity would be more complicated because several pulses have to be taken into account in the analytical expression of the filtering. 

The extrema of the beating amplitude are then expressed:

For $\Delta\phi(t,\tau)= 0\,\,(2k\pi)$,

\begin{equation}
\begin{split}
I_{max}(t,\tau) = &\,  E_i(t,0) \, E_i^*(t,0) + E_j(t,\tau) \, E_j^*(t,\tau)\\
                  &\;  + 2 \sqrt{ E_i(t,0) \, E_i^*(t,0) \, E_j(t,\tau) \, E_j^*(t,\tau)}    
\end{split}
\label{Eq.Equation2}
\end{equation}

For $\Delta\phi(t,\tau) = (2k+1)\pi$,

\begin{equation}
\begin{split}
I_{min}(t,\tau) = &\,  E_i(t,0) \, E_i^*(t,0) + E_j(t,\tau) \, E_j^*(t,\tau)\\
                  &\;  - 2 \sqrt{ E_i(t,0) \, E_i^*(t,0) \, E_j(t,\tau) \, E_j^*(t,\tau)}    
\end{split}
\label{Eq.Equation3}
\end{equation}

In order to determine the contrast of the beating, it is important to measure $I_{min} (t)$ and $I_{max}(t)$. The Histogram routine of the sampling oscilloscope allows this measurement from the information on the width and the voltage values of the histogram. Several considerations have to be done:\\
For a narrow histogram width, we consider the variable $t$ as a constant in the expressions~\ref{Eq.Equation2} and \ref{Eq.Equation3}. Moreover, $t$ is equal to zero at the top of the pulse. Then, if the position of the histogram window is the same all along the contrast measurement:
- The variables $t$ and $\tau$ are constant for each measurement
- The variable $t$ could be equal to zero and then neglected in the following expressions of the beating contrast $C$ at the maximum of intensity of the reference pulse becomes.

\begin{equation}
C (\tau)= \frac {I_{max}(\tau)-I_{min}(\tau)} {I_{max}(\tau)+I_{min}(\tau)} = \frac {2 \sqrt{ E_i\, E_i^* \, E_j(\tau) . E_2^*(\tau)}} {E_i \, E_i^* + E_j(\tau) \, E_j^*(\tau)}
\label{Eq.Equation4}
\end{equation}

The pulse shape $I_j(\tau)$ is directly related to the intensity ratio $R(\tau)$ between the two pulses in the histogram window as:

\begin{equation}
R (\tau)= \frac {E_j(\tau)\,E_j^*(\tau)} {E_i\,E_i^*1}  =  \frac{I_j(\tau)} {I_i} 
\label{Eq.Equation5}
\end{equation}

Finally, the relation between the beating contrast $C(\tau)$ and the pulse intensity $R(\tau)$.

\begin{equation}
R (\tau)= \left[ \left( \frac {1} {C(\tau)}  \right)- \left( \frac {1} {C^2(\tau)} -1 \right) ^{\frac{1}{2}} \right] ^2
\label{Eq.Equation6}
\end{equation}

By this way, we can determine the shape of $I_2$ with the resolution corresponding to the Optical Delay Line precision.

\subsection{Bandwidth photodetector limitations}

The dependency on the photodetector bandwidth of the BCM technique has been studied. The beating response depends intrinsecally to the detector response $h(t)$. The formalism used for this study is expressed as following:

\begin{equation}
I_F(t,\tau)=I(t,\tau) \ast h(t) = \int_{-\frac{T_H}{2}}^{+\frac{T_H}{2}} I(t,\tau) \, h(u-t)\, du 
\label{Eq.Equation7}
\end{equation}


In order to simplify the expression of the beating detected with a limited bandwidth photodetector, we consider the following hypothesis : 

- The contrast measurement is determined using a histogram window bounded by $\left[ T_H/2;T_H/2 \right]$.

- We consider that the position of the window coincides with the top of the reference pulse and this, for each measurement.
In a realistic approach, the histogram window width is variable and must be adjusted in order to optimize the measurement. We will discuss this point at the end of this section.

- We consider that for each value of delay $\tau$, the histogram is completed within the acquisition time $T_A$. This time is necessary for covering all the values of the phase difference between the pulses and produce a statistical histogram that make use for the contrast measurement. 

- For the contrast measurements, we only consider the convolution of the beating in the selected histogram window $T_H$. 
In the histogram window, the temporal response of filter $h(t)$ can be considered as a constant in order to simplify the expression of the convolution product which is contained in the global constant $K\,=\,T_A \times T_H$. For example, a 100~ps photodiode response which corresponds to a 10~GHz bandwidth ($B=10~GHz$) should be simplified by a constant if the measurement in realized in the histogram time window of few picoseconds. The condition for this hypothesis is that: $ T_H\,<<\,1/B $.

According to the distributivity of the convolution product and the temporal integration between the boundaries determined by the measurement time window, we give the expression of the three different terms. We notice that $I_i$ is the reference and does not depends on the delay $\tau$:

So, the expression of the intensity is composed of three terms $A_1$, $A_2$ and $A_{1,2}$ :

\begin{equation}
I_F(\tau)=\int_{-\frac{T_A}{2}}^{+\frac{T_A}{2}} I_F(t,\tau)\,dt = A_1(\tau) +  A_2(\tau) + A_{1,2}(\tau)     
\label{Eq.Equation9}
\end{equation}

\begin{equation}
\begin{split}
A_1(\tau) &\,= \int_{-\frac{T_A}{2}}^{+\frac{T_A}{2}} \left[ I_i(t)\ast h(t) \right]\,dt = 
\int_{-\frac{T_A}{2}}^{+\frac{T_A}{2}} \left[  \int_{-\frac{T_H}{2}}^{+\frac{T_H}{2}} I_i(t) \, h(u-t)\, du \right]\,dt\\
&\;= K \langle I_i(\tau) \rangle
\end{split}
\label{Eq.Equation10}
\end{equation}

\begin{equation}
\begin{split}
A_2(\tau) 	&\,= \int_{-\frac{T_A}{2}}^{+\frac{T_A}{2}} \left[ I_j(t,\tau)\ast h(t) \right]\,dt = 
	\int_{-\frac{T_A}{2}}^{+\frac{T_A}{2}} \left[  \int_{-\frac{T_H}{2}}^{+\frac{T_H}{2}}  I_j(t,\tau) \, h(u-t)\, du \right]\,dt\\
				&\;= K \langle I_j(\tau) \rangle    
\end{split}
\label{Eq.Equation11}
\end{equation}

\begin{equation}
\begin{split}
A_{1,2}(\tau) 	&\,= \int_{-\frac{T_A}{2}}^{+\frac{T_A}{2}} \left\{ \left[ 2\sqrt{I_i(t)\,I_j(t,\tau)}\cos\left[\Delta \phi_{chirp}^{i\neq j}(t) \right]\right]\ast h(t) \right\} \,dt \\
					&\;= K \langle 2\sqrt{I_1\,I_2(\tau)}\cos\left[\Delta \phi_{chirp}^{i\neq j}(t)\right] \rangle
\end{split}
\label{Eq.Equation12}     
\end{equation}

Also, the following expressions give the extrema of the filtered beating, deduced from Equations (\ref{Eq.Equation2}) and (\ref{Eq.Equation3}) with a distribution of the phase difference between 0 and 2$\pi$:

\begin{equation}
\begin{split}
I_F^{max}(\tau) = K [\langle I_1 \rangle + \langle I_2(\tau) \rangle + \langle 2\sqrt{I_1\,I_2(\tau)} \rangle]\\
I_F^{min}(\tau) = K [\langle I_1 \rangle + \langle I_2(\tau) \rangle - \langle 2\sqrt{I_1\,I_2(\tau)} \rangle]
\end{split}
\label{Eq.Equation13}    
\end{equation}

Therefore, we are able to write the expression of the beating contrast with an ideal detector. We show that the filtering effect is, according to the hypothesis, negligible:

\begin{equation}
\begin{split}
C_F(\tau)&\,= \frac{I_F^{max}(\tau)-I_F^{min}(\tau)}{I_F^{max}(\tau)+I_F^{min}(\tau)} =
 \frac{2K \langle 2\sqrt{I_1\,I_2(\tau)}\rangle}{2K\left\{ \langle I_1 \rangle + \langle I_2(\tau) \rangle \right\} }\\
C_F(\tau)&\,= C(\tau)
\end{split}
\label{Eq.Equation14}    
\end{equation}

Thus, thanks to the model hypothesis, the filtered beating contrast must be theoretically considered equal to the ideal beating contrast. According to this result, we neglect in the BCM measurement the filtering effect generated by the photodetector. Thus, the use of a low filter bandwidth allows the use of a wider histogram window. This point is really important because in the reality, larger is the histogram window is large and higher is the number of hits and faster is the measurement is fast. This discussion has been taken into account in the following of the study. The histogram window is fixed at 5 ps for all the simulations and experimental measurements.
The second point is that the detector bandwidth has to be higher than the repetition rate in order to maximize the contrast and then enhance the sensitivity of the measurement. This condition is one of the limitations of the measurement.
In the following part, we have done several simulations thanks to the theoretical model in order to estimate the benefits and the limitations of the technique.

\section{Modelling results}

The BCM technique modelling has been computed using the analytical expressions presented in the previous part. We present here different results obtained thanks to the equation~\ref{Eq.Equation14} which takes into account the realistic photodetector response. Most of the parameters which have been introduced in the computation are extracted from the experimental part of the study.

Three studies have been done:

\begin{itemize}
	\item{Impact of the bandwidth on the pulsewidth measurement}
	\item{Measurement of pulse pedestals and impact of the bandwidth on it}
	\item{Influence of the noise (optical and/or electrical noise) on the measurement of the pulse extinction ratio}
\end{itemize}

\begin{figure}[!ht]
\centering
\includegraphics[scale=1]{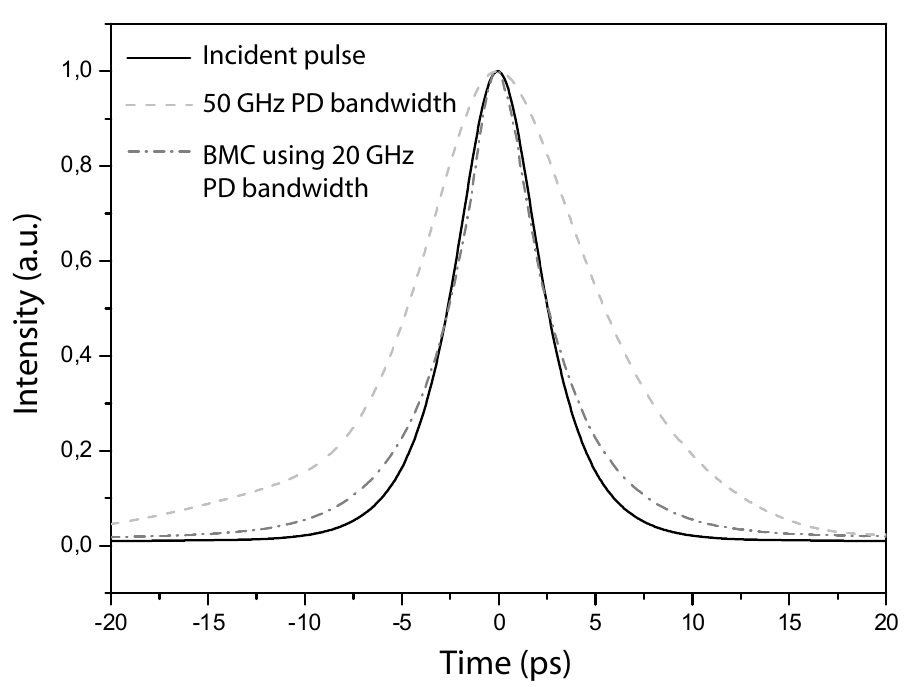}
\caption{Comparison of the BCM technique with direct detection using a 50~GHz bandwidth photodetector. The pulse to measure is 4.5~ps (continuous line), the detected pulse is 12.5~ps (dash line) and the pulse measure using BCM technique is 4.5~ps (dash and dot line).}
\label{fig.figure3}
\end{figure}

In order to validate the simulation approach, we present qualitative results obtained using the simulation. In Figure~\ref{fig.figure3}, we show the theoretical pulse to be characterized (continuous line). The input pulse has an hyperbolic secant shape with 4.5~ps FWHM. The dash curve is the theoretical response of a 50~GHz bandwidth photodetector. That provides a 12.5~ps pulse. The dot and dash curve shows the pulse trace obtained using the BCM technique which produces a 4.5~ps FWHM. The photodetector simulated for the BCM technique is 20~GHz bandwidth corresponding to a $2^{nd}$ order Butterworth filter function for $h(t)$. The width of the histogram window is 5~ps and the delay between the incident pulses is a fraction of the incident pulse-width. The sampling rate of the delay is chosen in order to optimize the time of calculation and the measurement resolution. Moreover, we introduce  a variable additional noise in the simulation in order to analyze the sensitivity of the technique to the signal to noise ratio (SNR) on the input signal. in these conditions, we observe in one hand, that the method is really efficient and on the other hand, that the hypothesis done in the theoretical part are confirmed.
In the following of the study, we present several results allowing us to determine the performances expected with the BCM technique.

\subsection{Pulsewidth analysis}

\begin{figure}[!ht]
\centering
\includegraphics[scale=1]{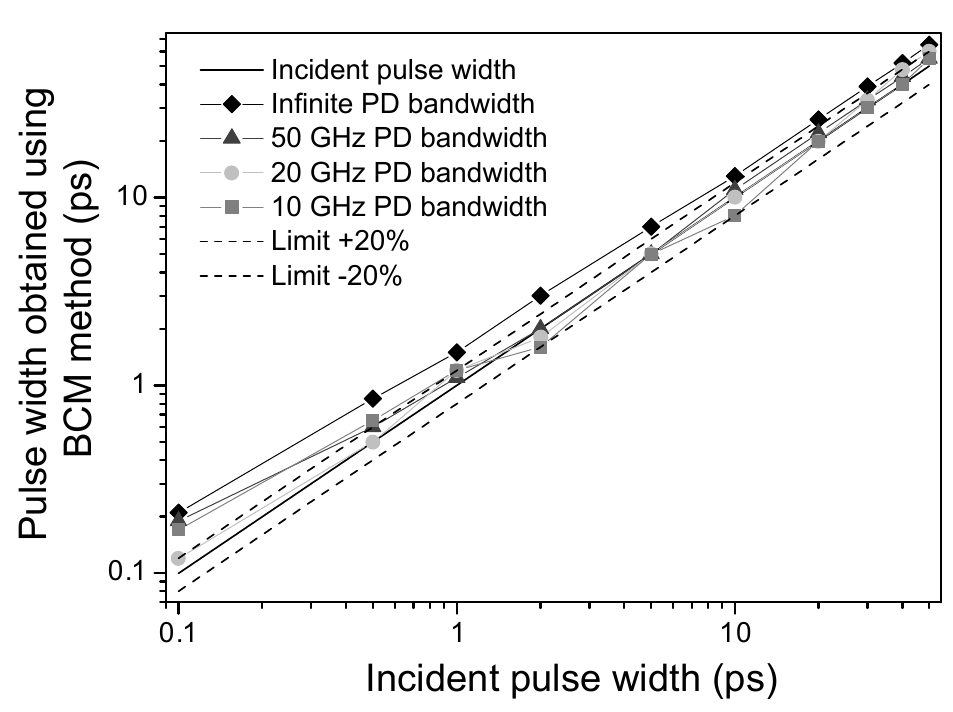}
\caption{Analyse of the efficiency of the BCM technique for the pulsewidth estimation. Results of the BCM are compared for different photodetector bandwidths: 10~GHz, (diamonds), 20~GHz (down triangles), 50~GHz (circles) and infinite bandwidth (squares). The dashed line gives the real width of the pulse.}
\label{fig.figure4}
\end{figure}

In this part we present the theoretical study concerning the pulsewidth measurement using the BCM technique. We test the calculation with different pulsewidths from 100~fs to 50~ps. We also modify the detector bandwidth used for the simulation in order to determine the most suitable one according to the previous hypothesis. Results are shown in Figure~\ref{fig.figure4}. The curve with down triangles corresponds to the 20~GHz bandwidth and is very close to the continuous line which represents the real pulsewidth. The result with this bandwidth gives very reliable results because each calculated  pulsewidths are included in the range corresponding +/- 20\% of the true width.
On the other hand, the infinite bandwidth does not give the best result because the corresponding temporal response is lower than the histogram width. This critical parameter leads us to choose the histogram window width in order to keep a flat response of the detector in the window and then to avoid nonlinearities in the measurement. We clearly observe on the simulation results that the bandwidth parameter introduces important variations on the pulsewidth measurement. This behaviour is linked to the approximation made on the photodetector response that imposes $h(t)$ as a constant.
Moreover we show that the 10~GHz photodetector gives us the most reliable pulsewidth measurement. 

Another consequence linked to the histogram window width, is the speed of the measurement which will be increased experimentally by a wider histogram window. This point is also important for the choice of this parameter especially for the experimental part.

In the next part, we focus our study on the measurement of the pulse shape obtained using the BCM technique and especially on the detection of the pedestals which are not detectable using classical methods as direct photo-detection and autocorrelation.

\subsection{Pulse shape analysis}

\begin{figure}[!ht]
\centering
\includegraphics[scale=1]{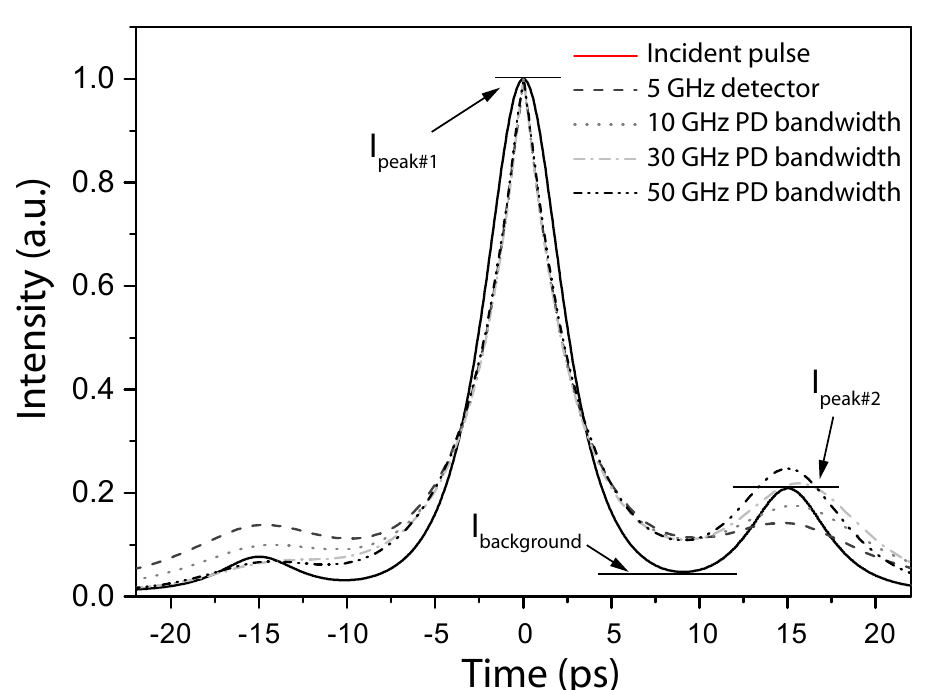}
\caption{Pedestal measurement using the BCM technique. Results of measurement for different photodetector bandwidths are compared to the incident pulse shape: 5~GHz (dashed line), 10~GHz (dotted line), 30~GHz (dash-dotted line) and 50~GHz (dash-dot-dotted line).}
\label{fig.figure5}
\end{figure}

We present a study on the capability of the BCM technique to pedestal detection. So we simulate two pedestals on each side of the pulse in order to determine the best configuration allowing the measurement of the pedestals position and amplitude.
Results are shown in Figure~\ref{fig.figure5}. The incident pulse is composed of three 4.5~ps width hyperbolic secants separated by 15~ps from the central peak. The repetition frequency of the pulse train is still 10~GHz. The ratio between the pedestals and the normalized central peak is 7~dB for the right peak and 12~dB for the left peak. We simulate the BCM technique for different photodetector bandwidths. The first observation concerns the pulsewidth which is not correctly determined in presence of pedestals. The second observation concerns the pedestal position which is determined with the BCM technique. The third observation concerns the influence of the photodetector bandwidth on the pedestal measurement efficiency. By the way, we simulate the technique for 5, 10, 30 and 50~GHz photodetector bandwidths. We show clearly that higher is the bandwidth, more efficient is the measurement. The first conclusion is that the technique allows pedestal detection. The second conclusion is that the contrast of the pedestal obtained using the BCM technique is reduced as compared to the incident pulse. For this analyse, we use two criteria in Figure~\ref{fig.figure5}, based on the peak and background intensities values:

\begin{itemize}
	\item the contrast $C\,=\,\left(I_{peak}-I_{background}\right)\,/\,I_{background}$,
	\item the Extinction Ratio $ER\,=\,\left(I_{peak\#1}-I_{background}\right)\,/\,\left(I_{peak\#2}-I_{background}\right)$.
\end{itemize}

For the 50~GHz bandwidth configuration, the contrast of the pulses (central peak and pedestals) is divided by two as compared to the incident pulse, which is the best result obtained with the simulation. On the other hand, the extinction ratio is not modified as compared to the incident pulse. So this interesting result shows that, with a suitable calibration, the BCM technique allows a realistic measurement of pulse pedestals.

In the following part, we focus our study on the theoretical sensitivity of the technique to optical and/or electrical noise in excess introduced in the measurement process.

\subsection{Noise influence in the BCM technique}

\begin{figure}[!ht]
\centering
\includegraphics[scale=1]{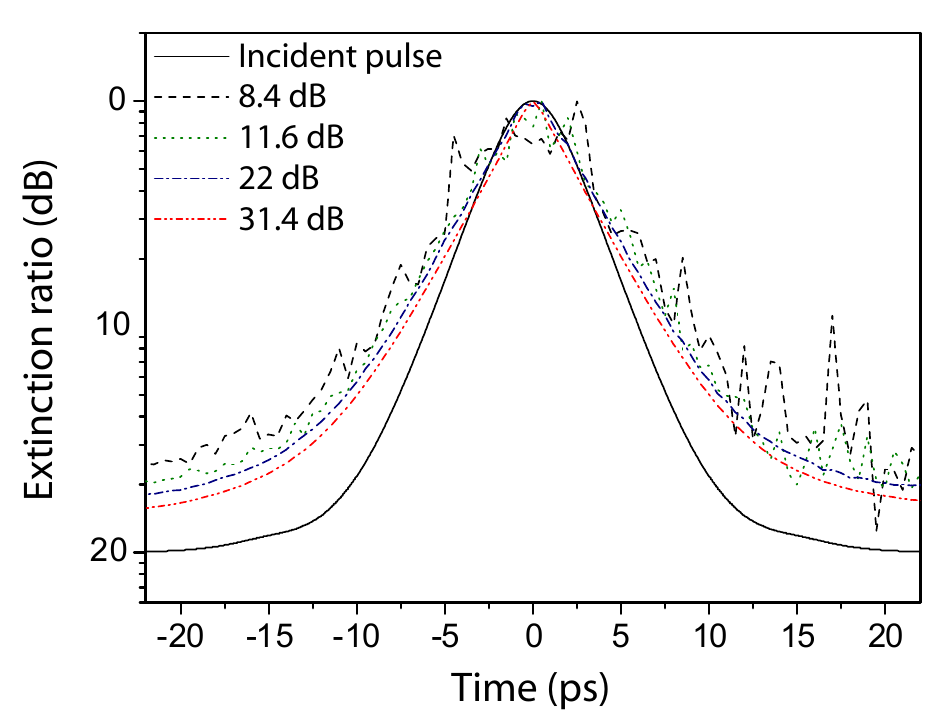}
\caption{Simulation of the BCM technique in presence of noise in the processus. The signal to noise ratio values are expressed in decibel for a signal bandwidth of 5~nm.}
\label{fig.figure6}
\end{figure}

Optical and electronic noises are considered as the main limitations in the photodetection process. Beatings between optical noise and signal are generated into the photodetector. Electronic resulting noise is proportional to the noise spectral density in the band $B$. In order to study the influence of the noise in the BCM technique, we simulated an additive white noise in a band B equal to 5~nm. The corresponding Signal to Noise Ratio (SNR) is expressed in dB/5~nm. The signal is still a pulse train at 10~GHz of repetition frequency with a 4.5~ps pulsewidth. The extinction ratio of the pulse train which is defined as the ratio between the maximum of the pulse and the background of the signal is fixed to 20~dB. The results are shown in Figure~\ref{fig.figure6}. We present the extinction ratio in a logarithm scale in order to observe precisely the performance of the BCM technique to determine the extinction ratio. So, we changed the SNR of the pulse train from 31.4~dB, which correspond to a noiseless signal, to 8.4~dB which corresponds to a degraded signal.  
The first observation is that the extinction ratio of the pulse train determined using the BCM technique is very close to the true value if the SNR is high. For example, the extinction ratio measured with a SNR of 31.4~dB is higher than 19~dB. We show that the reduction of the SNR generate a degradation of the extinction ratio measured using the BCM technique. The second observation is that the noise is transmitted through the BCM technique with a shape fluctuation corresponding to the sampling rate of the measurement (sampling rate of the delay line).
Also, the BCM technique is really sensitive to noise so, its minimisation is necessary for the method optimisation.  

\section{Experimental results and measurement validation}

\subsection{Experimental set-up}
The experimental set-up of the BCM technique has been already shown in Figure~\ref{fig.figure1}. For the experimental validation of the technique, we use a Mach-Zehnder interferometer composed of two single mode fibers (SMF) and 3~dB couplers. The length of the decorrelation fiber depends on the repetition rate of the optical source. The purpose of the measurement is the production of a beating between two decorrelated incident pulses. The decorrelation fiber length corresponds to a multiple of the repetition rate. The length also depends on the type of the pulse source, especially, on its coherency.
The all fibered motorized ODL has a typical resolution of 10~fs allowing to control precisely the delay between the pulses. The pre-amplified photodetector module has a bandwidth $B_o$=~20~GHz. The sampling oscilloscope has an electrical bandwidth limited to 40~GHz. The equivalent bandwidth is then close to 20~GHz.


According to the equations~\ref{Eq.Equation12} and \ref{Eq.Equation13}, the contrast $C_F(\tau)$ of the beating is experimentally measured using the histogram function of a \textit{CSA~803} sampling oscilloscope from \textit{Tektronix, Inc}. This measurement gives the statistical distribution of the beating voltage in the controlled measurement time window. A time window of $T_H$ = 5~ps is used in order to obtain a flat signal for the histogram measurement and then to confirm the hypothesis developed in the simulation part:($T_H\,<<\,1/B$).
The duration of the histogram measurement is about 5 seconds in order to store a great number of samples with all values for the phase difference between the two pulses. Then, we measure the mean value, the maximum and the minimum at 3$\sigma$ of the distribution corresponding to the routine for >99\% of shots. This experimental process allows the measurement of the contrast $C_F$ and then the obtaining of the ratio between the pulse shape and the reference $R$. For the calculation, we take into account the loss of the set-up and the ratio between the two pulse train. This calibration is necessary to optimize the BCM technique.

Using a stable experimental set-up and an optimized routine for the calculation, we present in the following part experimental results of the BCM technique with two different pulsed sources in order to validate the technique.

\subsection{Results and comparison with autocorrelation trace}

In this section, we wish to definitely validate the BCM technique using a comparison between results with a reliable characterization technique of optical pulses. So, each optical pulse source is first characterized using an autocorrelator FR-103MN from $Femtochrome$. Each result obtained with the BCM technique is auto-convoluated mathematically in order to be compared with the experimental autocorrelation trace.

\subsubsection{BCM validation using gain-switched pulse source}

\begin{figure}[!ht]
\centering
\includegraphics[scale=0.8]{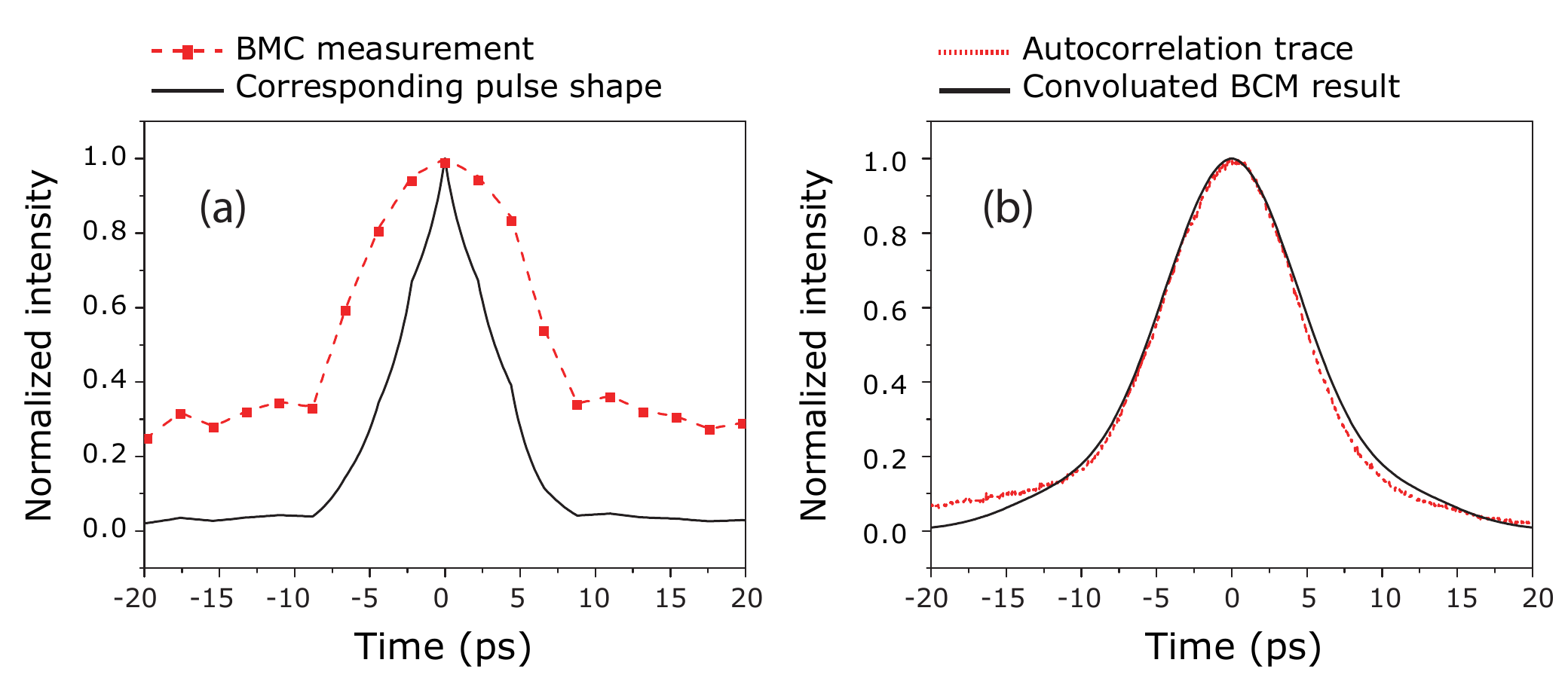}
\caption{First pulsed source characterization:  figure (a) presents the experimental results of the beating contrast and the calculated pulse shape; figure (b) shows the comparison between the autocorrelation trace and the auto-convolution trace of the shape obtained with the BCM technique.}
\label{fig.figure7}
\end{figure}

The first optical source is a 10~GHz internally modulated DFB laser from \textit{Alcatel OPTO+}~\cite{bouchouleSTQE1997}. The bias polarization current is 25~mA in order to generate a dynamic carrier response above 10~GHz. The 10~GHz electrical clock modulation is 16~dBm in order to obtain the shortest optical pulse of 23~ps at 10~GHz, characterized using a 45~GHz photodetector. The pulses generated by the laser are highly chirped by the internal current modulation. They are then compressed in 800~m of 29~ps/nm/km dispersion compensation fiber (DCF) in order to obtain the shorter pulses~\cite{chusseauOL1994}. An optimal pulse compression is then achieved.  In Figure~\ref{fig.figure7} (a), we present the results obtained using the BCM technique. We show firstly the measured beating contrast with a measurement resolution close to 1~ps. The calculation of the shape gives a pulsewidth of 6.3~ps with a symmetrical shape and an extinction ratio up to 20~dB. The pulse train is characterized using the autocorrelator. The autocorrelation trace which is shown in Figure~\ref{fig.figure7} (b), gives a convoluted width of 8.3~ps. We compared the result with the BCM result convoluated in Figure~\ref{fig.figure7} (b). We show firstly that the two traces are almost the same. Secondly,  the pulsewidth deducted from the autocorrelation trace deconvoluated with a Gaussian shape give a value for the pulsewidth of 5.8~ps which is very close to the BCM result.      
This repeatable experimental result shows that the BCM technique is validated for optical pulses with a compensated nonlinear chirp, generated by a Q-switched laser.

\subsubsection{BCM validation using mode-locked pulse source}

\begin{figure}[!ht]
\centering
\includegraphics[scale=0.8]{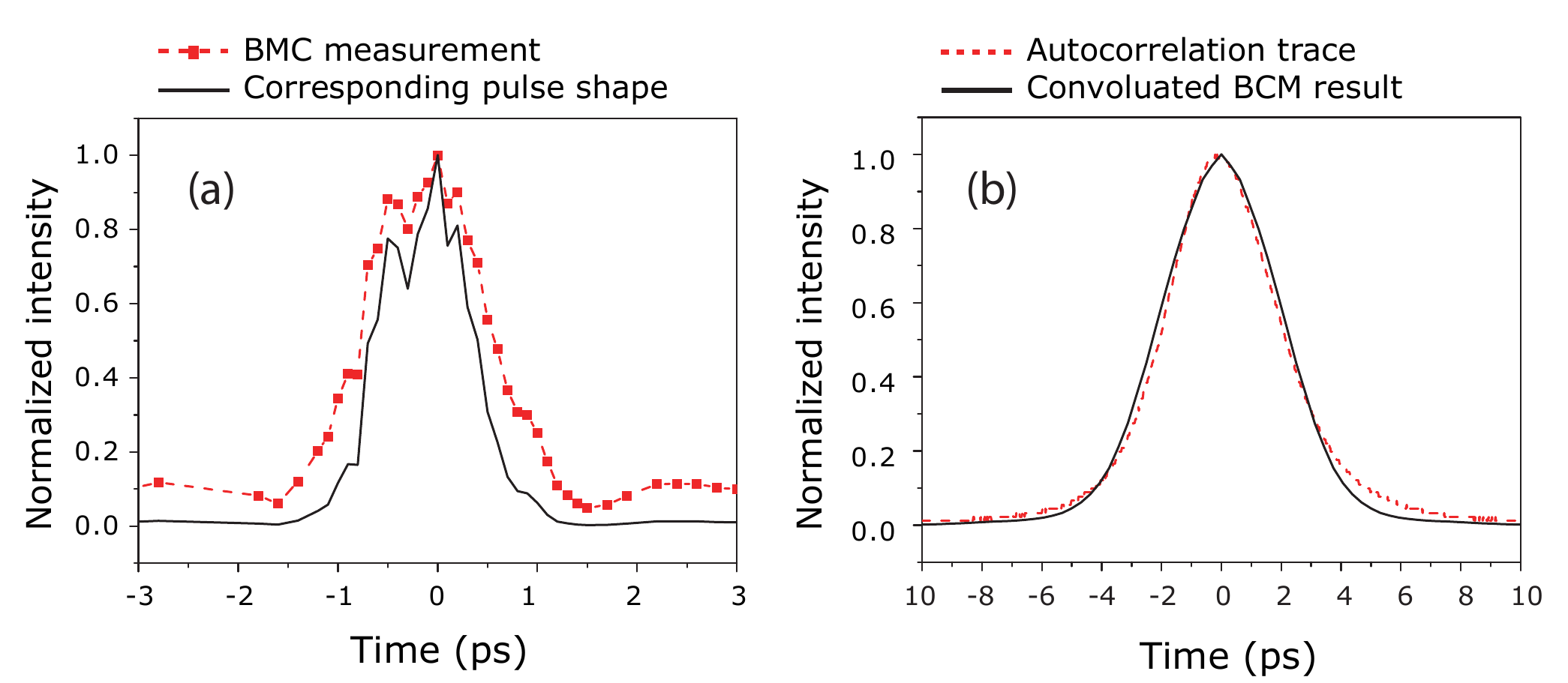}
\caption{Second pulsed source characterization :  figure (a) present the experimental results of the beating contrast and the calculated pulse shape; figure (b) shows the comparison between the autocorelation trace and the auto-convolution of the shape obtained with the BCM technique.}
\label{fig.figure8}
\end{figure}

The second optical source is a commercial erbium-doped fiber-based mode-lock Laser from \textit{Pritel,Inc}~\cite{BarnettOL1995} with 20~MHz of repetition frequency. The shortest optical pulses was obtained using an output 5~nm bandpass filter, centred at 1549~nm with a pump current of 150~mA. The constructor vouches for the Fourier transform limit condition of these optical pulses. The experimental characterizations are presented in Figure~\ref{fig.figure8}.

The BCM measurement is achieved with a high time resolution of 100~fs allowing a precise characterization of short pulses shown in Figure~\ref{fig.figure8} (a). We observe that the shape is not symmetrical and it does not look like a Gaussian shape. The pulsewidth measurement obtained using the BCM technique is 1.1~ps and the extinction ratio is higher than 20~dB. The auto-convolution of the trace gives a width of 4.4~ps. The comparison with the autocorrelation trace is shown in Figure~\ref{fig.figure8} (b). The autocorrelation trace gives a 4.2~ps pulsewidth. We observe that the two results are very close. This result validates once again the efficiency of the BCM technique for a non-chirped optical source.
An important point to be discussed is the reliability of the autocorrelator. In this experimental result, we found that the pulse shape is not Gaussian-like. With a Gaussian shape hypothesis, the deconvolution of the autocorrelation trace gives a pulsewidth value of 2.9~ps which is really far from the BCM technique result, while the auto-convoluated traces are very close. We understand with this result that the BCM technique does not depends on the initial pulse shape.

\subsection{Benefits and limitations of the BCM technique}
The method allows reliable pulse-width and pulse-shape measurement of short optical pulses with at least 100~fs of resolution. The asymmetry of the pulse shape is conserved. The contrast and the extinction ration of the pulse and the pedestals are obtained with a simple calibration of the experiment which depends principally on the photodetection block. The linear chirp of the optical pulses does not introduce any distortion on the measurement. Even if several parameters impose some limitations to the BCM technique, a suitable analysis of the experimental set-up must lead to an optimization of the measurement and give reliable results according to the pulse source type.  

\section{Conclusion}
Ultra-short optical pulses characterization using BCM technique offers wide new possibilities in terms of pulsewidth, contrast and pedestal extinction ratio measurements with high resolution. Simulation and experimental results have been presented in this study in order to validate and determine the limits of the technique. BCM measurement provides higher bandwidth than traditional photodetection, higher resolution than optical sampling equipment and more realist results than autocorrelators.
The technique uses commercial and standard optoelectronic devices specifically suitable for 10 GHz optical communication applications. The validation of the technique has been successfully done with optical sources used for 40~Gb/s and 160~Gb/s OTDM application and then, using a standard 20~Gb/s photodetector.

\section{Acknowledgments} This work was supported by the \textit{Ministère de la Recherche}, the \textit{Région\,Bretagne} and the European funding $FEDER$. The authors are especially grateful to Dr Blin for sharing his expertise and fruitful discussions on this new characterization method. We also want to thank Dr Joindot for its valuable assistance.

\bibliographystyle{IEEEtran}
\bibliography{IEEEabrv,biblioBCM}

\end{document}